\documentclass[epj,referee]{svjour}
\usepackage{graphics}

\begin{document}

\title{Local spin polarisation of electrons in Rashba semiconductor nanowires: Effects of the bound state}
\author{X.B. Xiao\inst{1,2}, F. Li\inst{1}, Y.G. Chen\inst{3}, \and N.H. Liu\inst{2,}
\thanks{nhliu@ncu.edu.cn}%
}                     

\institute{School of Computer, Jiangxi University of Traditional Chinese Medicine, Nanchang 330004, P.R. China. \and Institute for Advanced Study, Nanchang University, Nanchang 330031, P.R. China. \and Department of Physics, Tongji University, Shanghai 200092, P.R. China.}
\date{Received: date / Revised version: date}
%
\abstract{ The local spin polarisation (LSP) of electrons in two typical semiconductor nanowires under the modulation of Rashba spin-orbit interaction (SOI) is investigated theoretically. The influence of both the SOI- and structure-induced bound states on the LSP is taken into account via the spin-resolved lattice Green function method. It is discovered that high spin-density islands with alternative signs of polarisation are formed inside the nanowires due to the interaction between the bound states and the Rashba effective magnetic field. Further study shows that the spin-density islands caused by the structure-induced bound state exhibit a strong robustness against disorder. These findings may provide an efficient way to create local magnetic moments and store information in semiconductors.
} 
\authorrunning{X.B. Xiao et al.}
\titlerunning{Local spin polarisation of electrons in Rashba semiconductor nanowires: Effects of the bound state}

\maketitle

\section{Introduction}
The emerging field of semiconductor spintronics, i.e. trying to use electron spin rather than its charge to store and communicate information, has fueled tremendous research interest in recent years \cite{Zutic}. In the development of semiconductor spintronics, the Rashba SOI \cite{Rashba,Bychkov} plays a key role  since it is expected to coherently manipulate the electron spin state \cite{Datta} and its strength can be conveniently tuned by an external gate voltage \cite{Heida,Grundler,Nitta,Koga}. Subsequently, the effects of Rashba SOI in non-interacting \cite{Moroz} and interacting \cite{Hasler,Iucci,Malard} quantum wires have been investigated extensively because of their potential application for future spintronic devices.

As is well known, electrons in a nonmagnetic semiconductor are degenerate in spin state. Therefore, one of the main challenges in the development of semiconductor spintronics is to be capable of generating excess spin in semiconductor nanostructures, particularly by all electrical means. Various schemes have been proposed to satisfy
this goal such as spin injection \cite{Hanbicki}, the spin Hall effect (SHE) \cite{Murakami,Sinova} as well as spin filtering based on spin-polarised electron transport in two-terminal or multi-terminal semiconductor nanostructures in the presence of SOI(s) \cite{Yamamoto,Peeters,Wang,Zhang,Liu,Bellucci,Yamamoto2,Peeters2,Yokoyama,Japaridze}. Among these schemes, the SHE, which accumulates opposite spins at the two transverse edges of the two dimensional system due to the effects of SOI, has attracted increased interest \cite{Moca,Grimaldi,Shermana,Zyuzin,Malshukov,Lucignano,Wang2,Wang3,Chudnovsky,Silvestrov,Ma,Werake}. Recently, Brusheim and Xu have studied the LSP for electrons in a multichannel Rashba SOI-modulated straight waveguide. Besides the SHE, interesting Hall-like spin accumulations are also found in the internal region of the waveguide for a spin-unpolarised injection \cite{Brusheim}. This phenomenon stems from a spatial spin separation of coherent electron states. However, Rashba SOI applied to the system and the geometrical structure of the system may result in the localisation of electrons, and their effects on the LSP have not yet been considered in Ref. [38]. Further, the robustness of the LSP against disorder, which is essential for a real application, remains unclear. In our recent paper, we have investigated the spin-polarised electron transport properties of several typical Rashba semiconductor nanowires with longitudinal asymmetry and found that they are very sensitive to the bound states formed in the systems \cite{Xiao2}. Fano resonance or antiresonance structures have been found in the spin-dependent conductance in consequence of the bound states coupled to conducting states. In addition, the magnitudes of the spin polarisation around the Fano resonances or antiresonances are very large even in the presence of disorder.

Motivated by the two works above, in this paper, we study the LSP for electrons in two typical Rashba semiconductor nanowires with longitudinal-inversion symmetry when spin-unpolarized electrons are injected. Although the current in the output lead is still unpolarised, high spin-density islands with different polarities appear inside the nanowires as a result of the SOI- or structure-induced bound state. Furthermore, the spin-density islands caused by the SOI-induced bound state in a straight Rashba nanowire can be easily destroyed by disorder. However, the spin-density islands resulting from the structure-induced bound state in the wide-narrow-wide (WNW) Rashba nanowire can survive even in the presence of strong disorder, which demonstrates the feasibility of this structure for real information storage. The organisation of this paper is as follows. In Section 2, the theoretical model and the calculation method are presented. In Section 3, the numerical results are illustrated and discussed. A conclusion is given in Section 4.

\section{Model and analysis}
To study the problem we consider a two-dimensional electron gas (2DEG) formed in the $(x,y)$ plane of a semiconductor heterostructure. The 2DEG in the $(x, y)$ plane is restricted to a straight or WNW nanowire by a transverse confinement potential $V(x,y)$. The nanowire is subjected to an electrical field along the $z$-axis, which gives rise to a Rashba SOI. The strength of the Rashba SOI depends sensitively on the features of the quantum well, including the applied gate electric field \cite{Heida,Grundler,Nitta,Koga}, the ion distribution in the nearby doping layers \cite{Sherman2}, and the relative asymmetry of the electron density at the two quantum well interfaces \cite{Golub}. According to Ref. [39], Rashba nanowires of this type can be described by the spin-resolved discrete lattice model. The tight-binding Hamiltonian including the Rashba SOI on a square lattice is given as follows:
 \begin{eqnarray}
H=H_0+H_{so}+V,
\end{eqnarray}
where
\begin{eqnarray}
H_0=\sum\limits_{lm\sigma}\varepsilon_{lm\sigma}c_{lm\sigma}^{\dag}c_{lm\sigma}-t\sum\limits_{lm\sigma}\{c_{l+1m\sigma}^{\dag}c_{lm\sigma}\nonumber\\
+c_{lm+1\sigma}^{\dag}c_{lm\sigma}+H.c\},
\end{eqnarray}
\begin{eqnarray}
H_{so}=t_{so}\sum\limits_{lm\sigma\sigma'}\{c_{l+1m\sigma}^{\dag}(i\sigma_{y})_{\sigma\sigma'}c_{lm\sigma'}\nonumber\\
-c_{lm+1\sigma}^{\dag}(i\sigma_{x})_{\sigma\sigma'}c_{lm\sigma'}+H.c\},
\end{eqnarray}
and
\begin{eqnarray}
V=\sum\limits_{lm\sigma}v_{lm}c_{lm\sigma}^{\dag}c_{lm\sigma},
\end{eqnarray}
in which $c_{lm\sigma}^{\dag}(c_{lm\sigma})$ is the creation (annihilation) operator of an electron at site $(lm)$ with spin
$\sigma$, $\sigma_{x(y)}$ is the Pauli matrix, and $\varepsilon_{lm\sigma}=4t$ is the on-site energy with hopping
energy $t=\hbar^{2}/2m^{\ast}a^{2}$, where $m^{\ast}$ and $a$ are the effective mass of the electron and the lattice constant, respectively. $v_{lm}$ is the additional confining potential. The SOI strength is $t_{so}=\alpha/2a$, where $\alpha$ is the Rashba constant.

In ballistic transport, the spin-resolved conductance is given by the Landauer-B$\ddot{u}$ttiker \cite{Buttiker} formalism with the help of the non-equilibrium Green function formalism \cite{Pareek}. The two-terminal spin-resolved conductance for the Rashba nanowire is
\begin{eqnarray}
G^{\sigma'\sigma}=G_0Tr[\Gamma_{L}^{\sigma}G_{r}^{\sigma\sigma'}\Gamma_{R}^{\sigma'}G_{a}^{\sigma'\sigma}],
\end{eqnarray}
where $G_0=e^2/h$ is the conductance quantum, $\Gamma_{L(R)} =i[\sum_{L(R)}^{r}-\sum_{L(R)}^{a}]$, $\sum_{L(R)}^{r}=(\sum_{L(R)}^{a})^{\ast}$ is the self-energy from the left (right) lead, and $G_{r}^{\sigma\sigma'}(G_{a}^{\sigma'\sigma})$ is the retarded
(advanced) Green function of the Rashba wire including the impacts of the leads. The trace is over the spatial and spin degrees of freedom. The Green function above is computed by the spin-resolved recursive Green function method \cite{Xiao2}.

In the following calculation, all energies are normalised by the hopping energy $t(t=1)$, and the $z$-axis is chosen as the spin-quantised axis so that $|\uparrow>=(1,0)^{T}$ represents the spin-up state and $|\downarrow>=(0,1)^{T}$ denotes the spin-down state. For simplicity, the hard-wall confining potential approximation is
adopted to determine the boundary of the nanowires, since different confining potentials only change the positions of the subbands and the energy gaps between them. The total probability and the LSP of the $z$-component of electrons in the nanowires are described respectively as \cite{Brusheim,Zhai}:
\begin{eqnarray}
\rho^{total}=\sum\limits_{n,\sigma}\psi_{n\sigma}^{\dag}(\textbf{r})\psi_{n\sigma}(\textbf{r}),
\end{eqnarray}
and
\begin{eqnarray}
LP_{z}(\textbf{r})=\sum\limits_{n,\sigma}\psi_{n\sigma}^{\dag}(\textbf{r})\sigma_{z}\psi_{n\sigma}(\textbf{r}),
\end{eqnarray}
where $\psi_{n\sigma}(\textbf{r})$ is the velocity-normalised wave-function of the scattered state corresponding to electrons incident from the $n$th subband with spin $\sigma$ in the left lead, and the summation is taken over all the propagating modes in the lead.

\section{Results and discussion}
Figure 1(a) presents a sketch of a straight nanowire in the presence of Rashba SOI, connected to two leads with the same width. The Rashba wire has width 20a and length 50a. The two connecting leads are normal-conductor electrodes without SOI, since we are only interested in spin-unpolarised injection. The spin-dependent conductance for the straight Rashba nanowire as a function of the electron energy $E$ is shown in Fig. 1(b). The strength of the Rashba SOI is $t_{so}=0.08$. Step-like structures appear in the spin-dependent conductance when the electron energy $E>0.026$ because the subbands of the nanowire become propagating modes one by one \cite{Xiao2}. Apart from the step-like structures, oscillations also emerge in the spin-dependent conductance, resulting from the interference between the forward and backward electron waves caused by the SOI-induced potential well. Furthermore, it is worth noting that dip-like structures [see the (red) circle] emerge in the spin-dependent conductance when the electron energy approaches the thresholds of the propagating modes. This effect is attributed to the SOI-induced bound states formed in the quantum wire coupled to the conducting states through Rashba intersubband mixing, giving rise to pronounced Fano resonances \cite{Sanchez}. Due to the longitudinal inversion symmetry in the straight nanowire, the spin-up conductance ($G^{up}=G^{\uparrow\uparrow}+G^{\uparrow\downarrow}$) is always equals to the spin-down conductance ($G^{down}=G^{\downarrow\downarrow}+G^{\downarrow\uparrow}$). Figure 1(c) shows the total probability distribution $\rho^{total}$ of electrons in the straight Rashba nanowire. For electron energy $E=0.15$, at which the conductance is on the plateau, the total probability distribution displays two regular stripes that represent two propagating modes. According to Fig. 1(b), it can be concluded that electrons in the output lead are still unpolarised. However, the LSP of electrons inside the Rashba wire $LP_{z}$ is non-zero. As shown in Fig. 1(d), four regular spin-polarised stripes with alternating signs of polarisation emerge in the LSP distribution. The underlying physics stems from a spatial spin separation of coherent electron states in the nanowire \cite{Brusheim}. In contrast, for electron energy $E=0.218$, at which the conductance is at the dip, the regular stripe-like distributions of the total probability and LSP are disturbed. The total probability of electrons in the Rashba wire is very large in some places, with two obvious charge-density islands formed in the nanowire [see Fig. 1(e)]. Interestingly, the magnitudes of the LSP around these two charge-density islands are very large too, as depicted in Fig. 1(f), where two spin-density islands emerge in the wire.  This effect may result from the bound states interacting with the SOI-induced effective magnetic field, leading to the formation of the spin-density islands. Further, owing to the fact that the straight Rashba wire is symmetrical in the longitudinal direction, the two spin-density islands show different polarities.

Figure 2(a) presents a sketch of a Rashba SOI-modulated WNW nanowire. The narrow part of the wire has width $D=10a$ and length $H=30a$. The wide parts of the wire have the same width $W=20a$ and length $L=10a$, connected to two ideal normal-conductor leads with the same width W at each end. The spin-dependent conductance for the WNW Rashba nanowire as a function of the electron energy $E$ is shown in Fig. 2(b). The Rashba SOI strength is $t_{so}=0.153$. The step-like structure and oscillations exist in the spin-dependent conductance as seen in Fig. 1(b). However, a valley-shaped structure emerges on the conductance plateau [see the (red) circle]. This structure originates from coupling between the bound state caused by the nonuniform width and the  conducting states, resulting in a structure-induced Fano resonance. Figure 2(c) illustrates the total probability distribution of electrons in the WNW wire. The electron energy $E=0.220$, so that the conductance is in the valley. In contrast to the total probability distribution of electrons in the straight Rashba wire that spreads along the whole system, $\rho^{total}$ is strongly localised in the wide region to the right, i.e. two very high charge-density islands are formed in this region. The formation of these two charge-density islands originates from the interference between the forward electron waves and the backward electron waves reflected from the right wire-lead interface. As a consequence, there is a certain probability that electrons will stay in the wide region \cite{Xiao2}. Surprisingly, the magnitudes of the LSP around the two charge-density islands are also very large, which can be seen in Fig. 2(d), where four spin-density islands with alternating signs of polarisation appear in the wide region to the right. The physical mechanism associated with these spin-density islands is the same as that in Fig. 1(f).

The above calculation assumes a perfectly clean system, where there is no elastic or inelastic scattering. However, in a realistic system, there will be many impurities in the sample. Consequently, the effect of disorder on the LSP should be considered in practical applications. The effects of impurities in the sample can be introduced by fluctuation of the on-site energies, which are distributed randomly within a range of width $w$: $\varepsilon_{lm\sigma}= \varepsilon_{lm\sigma}+w_{lm}$, here $-w/2<w_{lm}<w/2$. In addition, the effects of impurities in the sample can also be incorporated via the Rashba SOI by allowing the strength of the Rashba SOI $t_{so}$ to fluctuate randomly in space, giving rise to many new phenomena such as the realisation of the minimum possible strength of SOI \cite{Sherman2} and the localisation of the edge electrons for sufficiently strong electron-electron interactions \cite{Strom}. For simplicity, in the present paper only the fluctuation of the on-site energies is considered, while that of the Rashba SOI strength is neglected. The LSP distribution of electrons in the two Rashba nanowires with the introduction of strong disorder are illustrated in Fig. 3. Except for the disorder strength $w=0.6$, the other parameters in Fig. 3(a) and (b) are the same as those in Fig. 1(f) and 2(d), respectively. The magnitude of the LSP of electrons in the straight Rashba wire [Fig. 3(a)], in contrast to that in Fig. 1(f), is increased or decreased by the disorder. In Particular, the spin-density islands in the straight Rashba wire are destroyed by the disorder. However, for the disordered WNW Rashba wire, as plotted in Fig. 3(b), the salient features of the LSP distribution are the same as those in Fig. 2(d). Further, the four spin-density islands can survive even when strong disorder is present.

\section{Conclusion}
In conclusion, we have studied the LSP of electrons in Rashba SOI-modulated semiconductor nanowires for a spin-unpolarised electron injection. High spin-density islands with alternating polarisation directions are formed in the nanowires by virtue of the SOI- or structure-induced bound states. Furthermore, the spin-density islands in the WNW Rashba wire caused by the structure-induced bound states can survive even in the presence of strong disorder. These results indicate that the proposed nanowires can be utilised to create local magnetic moments and store information without applying an external magnetic field.

\section*{}
This work was supported by the National Natural Science Foundation of China under Grant No. 10832005, 10774112 and the China Postdoctoral Science Foundation under Grant No. 20100481188.

%

\newpage

\begin{figure}
\resizebox{1\columnwidth}{!}{%
  \includegraphics{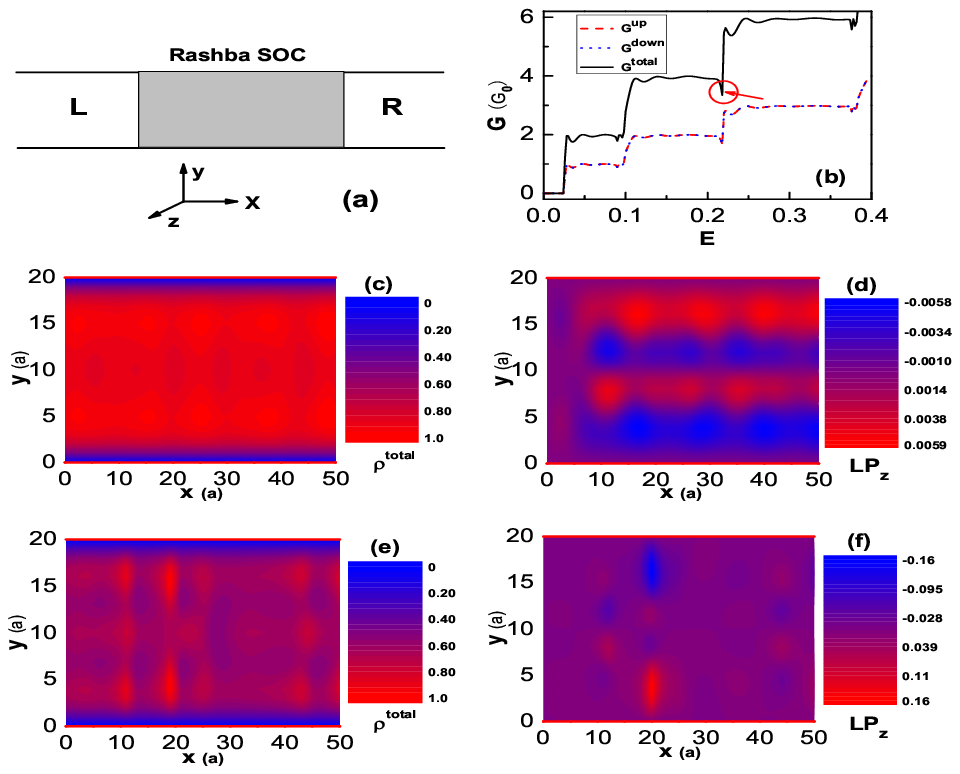}
  }
\caption{(Colour online) (a) Schematic diagram of the straight Rashba nanowire, connected to two leads with a vanishing SOI. (b) The calculated spin-dependent conductance as a function of the electron energy. The total probability distribution $\rho^{total}$ [(c) and (e)] and the corresponding LSP distribution [(d) and (f)] for electrons in the straight Rashba nanowire for a spin-unpolarised injection. The Rashba SOI strength is $t_{so}=0.08$. The electron energy is taken as $E=0.15$ in the middle panels and 0.218 in the lower panels.}
\label{fig:1}       
\end{figure}
\begin{figure}
\resizebox{1\columnwidth}{!}{%
  \includegraphics{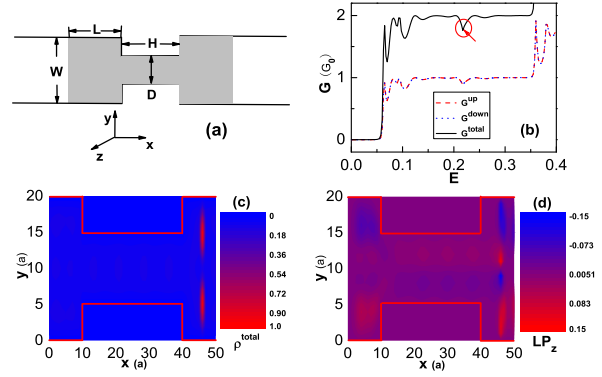}
  }
\caption{(Colour online) (a) Schematic diagram of the WNW Rashba nanowire, connected to two leads with a vanishing SOI. (b) The calculated spin-dependent conductance as a function of the electron energy. The total probability distribution (c) and the corresponding LSP distribution (d) for electrons in the WNW Rashba nanowire for a spin-unpolarised injection. The Rashba SOI strength is $t_{so}=0.153$ and the electron energy is taken as $E=0.220$.}
\label{fig:2}       
\end{figure}

\begin{figure}
\resizebox{1.0\columnwidth}{!}{%
  \includegraphics{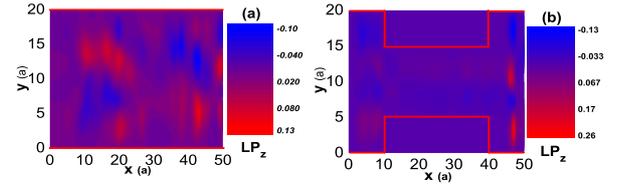}
 }
\caption{(Colour online) The LSP distribution for electrons in the disordered straight (a) and WNW (b) Rashba nanowires. The strength of disorder is $w=0.6$. The other parameters in (a) and (b) are the same as those in Figs. 1(f) and 2(d), respectively.}
\label{fig:3}       
\end{figure}

\end{document}